\documentclass[%
 reprint,superscriptaddress,
 amsmath,amssymb,
 aps,
prb,
]{revtex4-2}

\usepackage{graphicx}
\usepackage{dcolumn}
\usepackage{bm}
\usepackage[pdftex,colorlinks=true,citecolor=blue,linkcolor=blue,urlcolor=blue]{hyperref}
\hypersetup{pdfpagemode=UseNone,pdfpagelayout=OneColumn,pdfstartview=FitH}

\begin{document}


\title{
Search for magnetoacoustic quantum oscillations in the insulating phase
of \texorpdfstring{YbB$_{12}$}{YbB12}
}

\author{Ryosuke Kurihara}
\affiliation{Department of Physics and Astronomy, Tokyo University of Science,
Noda, Chiba 278-8510}
\email{r.kurihara@rs.tus.ac.jp}

\author{Atsuhiko Miyata}
\affiliation{Institute for Solid State Physics, The University of Tokyo,
Kashiwa, Chiba 277-8581, Japan}

\author{Koji Araki}
\affiliation{Department of Applied Physics, National Defense Academy,
Yokosuka, Kanagawa 239-8686, Japan}

\author{Shusaku Imajo}
\affiliation{Department of Advanced Materials Science, University of Tokyo,
Chiba 277-8561, Japan}

\author{Ruo Hibino}
\affiliation{Department of Physics, Kobe University, Kobe, Hyogo 657-8501, Japan}

\author{Atsushi Miyake}
\affiliation{Institute for Materials Research, Tohoku University, Oarai,
Ibaraki 311-1313, Japan}

\author{Sergei Zherlitsyn}
\affiliation{Hochfeld-Magnetlabor Dresden (HLD-EMFL) and W\"urzburg-Dresden
Cluster of Excellence ctd.qmat, Helmholtz-Zentrum
Dresden-Rossendorf, 01328 Dresden, Germany}

\author{Joachim Wosnitza}
\affiliation{Hochfeld-Magnetlabor Dresden (HLD-EMFL) and W\"urzburg-Dresden
Cluster of Excellence ctd.qmat, Helmholtz-Zentrum
Dresden-Rossendorf, 01328 Dresden, Germany}
\affiliation{Institut f\"ur Festk\"orper- und Materialphysik,
Technische Universit\"at Dresden, 01062 Dresden, Germany}

\author{Hiroshi Yaguchi}
\affiliation{Department of Physics and Astronomy, Tokyo University of Science,
Noda, Chiba 278-8510}

\author{Fumitoshi Iga}
\affiliation{College of Science, Ibaraki University, Mito 310-8512, Japan}

\author{Masashi Tokunaga}
\affiliation{Institute for Solid State Physics, The University of Tokyo,
Kashiwa, Chiba 277-8581, Japan}

\author{Yasuhiro H. Matsuda}
\affiliation{Institute for Solid State Physics, The University of Tokyo,
Kashiwa, Chiba 277-8581, Japan}

\date{\today}

\begin{abstract}
A highly exotic phenomenon in solid-state physics is the observation of magnetic quantum oscillations in insulators.
For instance, in the Kondo insulator YbB$_{12}$ various groups reported the observation of such oscillations seemingly originating from Fermi surfaces, though this contradicts the concept of an insulator having no charged quasiparticles. 
In this study, we searched for quantum oscillations in YbB$_{12}$ by using bulk-sensitive ultrasonic experiments in high magnetic fields up to 65 T and down to 485 mK.
For that, we utilized an YbB$_{12}$ single crystal that, in previous experiments, revealed oscillations in the magnetoresistance in the insulating state.
We confirmed oscillation-like behavior of the magnetoresistance as well as field-dependent oscillations in the magnetocaloric effect.
However, we could not observe magnetoacoustic quantum oscillations in the insulating state, only in the field-induced metallic state.
In the insulating state, we found some anomalies in our ultrasound data, the origin of which remains elusive. 
Our findings provide further information on the puzzling behavior of the insulating state of YbB$_{12}$.
\end{abstract}

\maketitle


\section{Introduction}
\label{sect_intro}

Magnetic quantum oscillations (MQOs) are important phenomena that serve to study the electronic properties of fermionic quasiparticles in solid-state physics.
Due to the formation of Landau levels in magnetic fields, the density of states (DOS) at the Fermi energy changes as a function of magnetic field. Several physical quantities, including the magnetization, magnetoresistance (MR), specific heat, and elastic constants, oscillate
periodically in the inverse magnetic field
\cite{Shoenberg_Text}.
In metals and semimetals, the frequency of these oscillations depends on the cross-sectional area of the Fermi surface (FS) perpendicular to the magnetic field.
This allows us to determine the FS topology of, for instance, conventional metals, heavy-fermion systems, and unconventional superconductors
\cite{Joseph_PR138, Joseph_PR148, Larson_PR156, Hall_PRB64, Mackenzie_PRL76, Analytis_PRB80}.
In addition, the observation of MQOs served for detecting metallic surface states in topological insulators
\cite{Ren_PRB82, Taskin_PRL107}.
Thus, the existence of itinerant fermionic quasiparticles forming a FS is
a key ingredient for the appearance of MQOs.

In contrast to metals, MQOs are not expected in insulators due to the absence of a FS.
Nevertheless, some groups recently reported the appearance of MQOs in insulators.
The first such report was for the Kondo insulator SmB$_6$ 
\cite{Li_Science346, Tan_Science349}.
In this material, the hybridization between the $4f$ and $5d$ electrons causes the appearance of an energy gap at the Fermi energy, resulting in insulating behavior at low temperatures
\cite{Fisk_PhysB223}.
The origin of these oscillations is strongly debated
\cite{Thomas_PRL122}.
Indeed, further careful investigations searching for MQOs in insulators are needed.
This may help to clarify whether these oscillations originate from a bulk Fermi surface, impurities, surface states, or from other exotic mechanisms.

\begin{figure*}[t]
\begin{center}
\includegraphics[clip, width=0.98\textwidth]{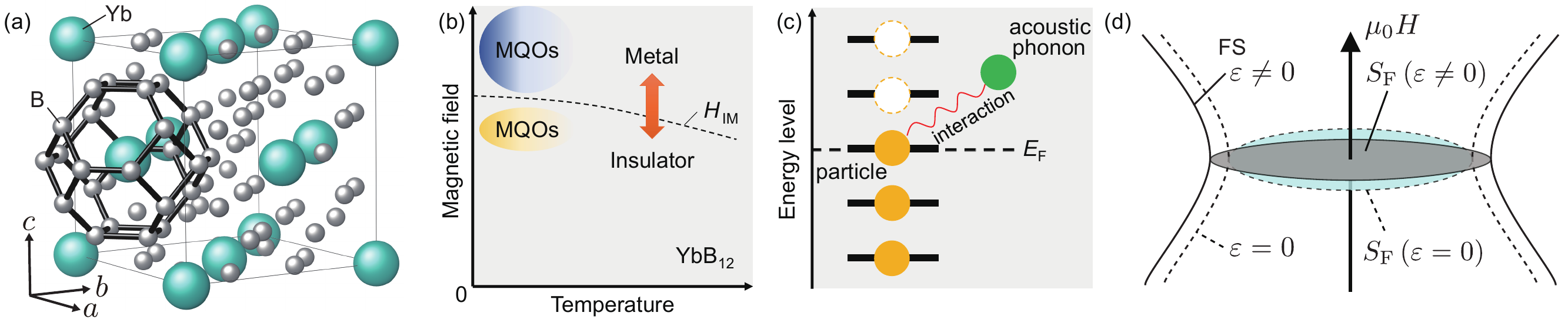}
\end{center}
\caption{
(a) Crystal structure of YbB$_{12}$ produced using \textsc{vesta} \cite{VESTA}.
(b) Sketch of the phase diagram of YbB$_{12}$.
At low temperatures, besides the observation of magnetic quantum oscillations (MQOs) in the
field-induced metallic phase, MQOs have been reported to appear in the insulating phase just below the insulator-metal (IM) transition field $H_\mathrm{IM}$.
(c) Sketch of the electron-phonon interaction leading to magnetoacoustic quantum oscillations. Particles at the Fermi energy ($E_\mathrm{F}$) interact with low-energy acoustic phonons with small wavevectors.
(d) Schematic illustration of the modulation of a Fermi surface (FS) and its extremal cross-sectional area $S_\mathrm{F}$, taken perpendicular to the magnetic field $\mu_0 H$, due to strain, $\varepsilon$, which is induced by the ultrasonic wave.
}
\label{Fig1}
\end{figure*}

Besides SmB$_6$, MQOs in the insulating state have been reported for YbB$_{12}$, another Kondo insulator
\cite{Liu_JPCM30, Xiang_Science362, Sato_NatPhys15}.
YbB$_{12}$ has the UB$_{12}$-type crystal structure with $Fm\overline{3}m$ ($O_h^5$) space group, displayed in Fig.\ \ref{Fig1}(a) 
\cite{Kasaya_JMMM31}.
The Yb ion is surrounded by a highly isotropic B$_{24}$ cage.
Several studies have revealed the existence of an energy gap at low temperatures
\cite{Kasaya_JMMM47, Iga_JMMM177, Iga_JMMM76, Takeda_PRB73}. 
Hybridization between the $5d$ and $4f$ electrons has been proposed as a mechanism for the band-gap opening
\cite{Saso_JPSJ72, Ohashi_PRB70}.
Despite this energy gap, various MR and magnetic-torque measurements have revealed MQOs in the insulating phase
\cite{Liu_JPCM30, Xiang_Science362, Sato_NatPhys15,miz24}
just below a field-induced insulator-metal (IM) transition at approximately 45 T
\cite{Sugiyama_JPSJ57, Iga_JPCS200, Terashima_JPSJ86, Terashima_PRL120}.
Above this field, different MQOs appeared in the field-induced metallic (FIM) phase
\cite{Xiang_NatPhys17, Liu_npj7},
as sketched in Fig.\ \ref{Fig1}(b).

One candidate mechanism for the appearance of MQOs in the insulating phase is the existence of surface states due to topological effects in YbB$_{12}$
\cite{Weng_PRL112, Hagiwara_Natcommun7, Sato_JPDAP54}.
However, the bulk-like FSs revealed by MR and magnetic-torque measurements, the distinct reduction of the oscillation amplitude in micro-structured samples, and the oscillations found in the heat capacity imply that these oscillations originate from the bulk
\cite{Xiang_Science362, Sato_JPDAP54,Chen_PRL135}.
Exotic quasiparticles with neutral charge have been proposed as a new mechanism for the MQOs
\cite{Sato_NatPhys15, Xiang_NatPhys17,Chen_PRL135}.
On the other hand, recent magnetocaloric-effect (MCE) and specific-heat measurements revealed anomalies below 35 T, indicating the existence of in-gap fermionic quasiparticle states \cite{Yang_NatCommun15}.
This unclear picture calls for further investigations on the origin of the observed oscillations, including a verification of whether they are genuine quantum oscillations.

In this work, we focused on the search for magnetoacoustic quantum oscillations (MAQOs) in YbB$_{12}$ by using ultrasonic measurements 
\cite{Shoenberg_Text, Luthi_Text}.
Figure \ref{Fig1}(c) shows a sketch of the involved electron-phonon interaction.
Due to their low energies (a typical frequency of $10^7$--$10^8$ Hz) and small wavevectors (a wavelength of $10^{-4}$--$10^{-6}$ m), ultrasonic waves are sensitive only to fermionic quasiparticles near the Fermi level, $E_\mathrm{F}$, that couple to acoustic phonons
\cite{Abrikosov_Text}.
The coupling is proportional to the number of quasiparticles at $E_\mathrm{F}$ and, therefore, to the DOS. 
With a changing magnetic field, the DOS oscillates according to the Landau-level quantization, which leads to MAQOs. 
From a bulk point of view, the FS is modulated by strain, $\varepsilon$, that is introduced into the crystal by the ultrasonic waves [Fig.\ \ref{Fig1}(d)].
This results in MAQOs in the elastic constant, which are proportional to the square of $ \partial \left( \ln S_\mathrm{F}  \right)/ \partial \varepsilon$, where $S_\mathrm{F}$ is an extremal cross-sectional area of the FS.
This bulk-sensitive technique is a powerful tool for Fermi-surface studies of, for example, heavy-fermion systems, semimetals, superconductors, and, recently, topological materials
\cite{Settai_JPSJ61, Yanagisawa_PRL123, Mavroides_PRL9, Toxen_PR137, Thompson_PRB4, Laliberte_PRB102, Nossler_PRB95, Scindler_PRB102, Ehmcke_PRB104}.

In a previous ultrasound experiment, we were not able to resolve MAQOs in YbB$_{12}$
\cite{Kurihara_PRB103}.
Since this might have been caused by an insufficient crystal quality, we decided to investigate the same sample that revealed MQOs in previous MR and magnetic-torque measurements
\cite{Xiang_Science362, Sato_NatPhys15}.
Using this sample, we measured elastic constants at high magnetic fields using ultrasound to search for MAQOs in the insulating phase of YbB$_{12}$.

This paper is organized as follows. In Sec.\ \ref{sect_Exp}, we introduce experimental details of sample preparation and ultrasonic measurements in pulsed magnetic fields.
In Sec.\ \ref{sub_characterizations}, we first present results of the characterization of the YbB$_{12}$ sample using high-field MR and MCE measurements.
As previously reported
\cite{Xiang_Science362, Sato_NatPhys15,Yang_NatCommun15},
we confirmed that the MR and MCE exhibit oscillation-like field dependence in the insulating phase.
In Sec.\ \ref{sub_ultrasound}, we show the results of our ultrasonic measurements. Although we observe some anomalies in the elastic constants $C_{11}$ and $C_{44}$, we again were not able to resolve quantum oscillations in the insulating phase of YbB$_{12}$.
In the field-induced metallic phase, on the other hand, we could observe MAQOs.
In Sec.\ \ref{sect_Discussion}, we discuss possible reasons why the MAQOs are absent in ultrasound measurements in the insulating phase.
Finally, we summarize our results in Sec.\ \ref{sect_Conclusion}.

\section{Experimental}
\label{sect_Exp}

The YbB$_{12}$ single crystal was grown using the floating-zone method
\cite{Iga_JMMM177}.
The crystal is identical to sample N3 of Ref.
\cite{Xiang_Science362}
and No. 1 of Ref. \cite{Sato_NatPhys15}.
We cut the sample into two pieces because of size limitations in our pulsed-field magnets.
We used Laue x-ray backscattering to determine the crystallographic orientation.
We aligned and polished the sample, resulting in six cubic (100) surfaces.
The size of the measured sample was $\sim 0.5$ mm $\times \sim 0.5$ mm $\times$ 3.631 mm.

We measured the MR of YbB$_{12}$ using a standard four-point method.
Gold wires fixed with silver paint (DuPont, 4922N) were used to contact the sample.
We utilized a numerical lock-in technique for resolving the alternating-current (AC) signals using a digital-storage oscilloscope and a function generator.
We applied AC signals with a frequency of 20 kHz and an excitation voltage of 300 mV$_\mathrm{pp}$ to the sample.
The resulting applied current was approximately 0.3--0.9 mA.

We performed quasi-adiabatic MCE measurements at an initial temperature of 0.6 K in pulsed magnetic fields up to 41 T
\cite{Kimura_PRB105, Imajo_RSI92, Imajo_JPSJ88}.
The sample was cooled using a $^3$He cryostat.
We monitored the temperature of the sample using a home-made RuO$_2$ resistive thermometer
deposited on a TiO$_2$ substrate and fixed to the YbB$_{12}$ sample using a small amount of Apiezon-N grease (ALLIANCE Biosystems).
We measured the resistance of the thermometer using a numerical lock-in technique with an AC excitation of 10 mV at a frequency of 10 kHz.

We used an ultrasonic pulse-echo method with a numerical vector-type phase-detection technique to measure the ultrasound velocities $v_{11}$ and $v_{44}$
\cite{Fujita_JPSJ80, Kohama_JAP132}.
We determined the longitudinal (transverse) elastic constant $C_{11} = \rho v_{11}^2$
($C_{44} = \rho v_{44}^2$) from $v_{11}$ ($v_{44}$) using the mass density $\rho = 4.828$ g/cm$^3$ and the lattice constant $a = 7.469$ \AA\ \cite{Kasaya_JMMM31}.
We employed piezoelectric LiNbO$_3$ transducers with a 36$^\circ$ Y-cut (41$^\circ$ X-cut)
(BOSTON PIEZO OPTICS INC. and Yamaju Ceramics Co.).
With these, we generated longitudinal (transverse) ultrasonic waves with fundamental
frequencies of about 22 (18) MHz for measurements at the Dresden High Magnetic Field Laboratory (HLD) and 30 MHz at the Institute for Solid State Physics (ISSP), the University of Tokyo. 
We used as well higher harmonic frequencies of 80 and 100 MHz (104 MHz), respectively, to obtain high-resolution data.
We glued the LiNbO$_3$ plates with room-temperature vulcanizing rubber (Toray Fine Chemicals and KE-42T, Shin-Etsu Silicone) or polysulfide polymer (Thiokol LP-32) to the sample.

As listed in Table \ref{Table_magnet}, we used three different nondestructive pulsed magnets
\cite{Miyata_IEEE36,zhe10}.
The pulse magnets were equipped with $^3$He cryostats.

\begin{table}[t]
\caption{
Specifications of the pulsed magnets at HLD and ISSP.
$\mu_0 H_\mathrm{max}$, $T_\mathrm{d}$, and $T_\mathrm{cool}$ indicate the maximum field, the duration of the pulses, and the cooling time of the magnet, respectively.
US, MR, and MCE stand for ultrasonic, magnetoresistance, and magnetocaloric effect, respectively.
}
\begin{ruledtabular}
\label{Table_magnet}
\begin{tabular}{ccccc}
Facility &$\mu_0 H_\mathrm{max}$ (T) & $T_\mathrm{d}$ (ms)  & $T_\mathrm{cool}$ (min)   &technique \\
\hline
HLD   &65  &150   &180  & US  \\
ISSP  &60  &36    &35   & US, MR  \\
ISSP  &41  &2000  &150  & MCE  \\
\end{tabular}
\end{ruledtabular}
\end{table}

\section{Results}
\label{sect_Result}

\subsection{Magnetoresistance and magnetocaloric effect}
\label{sub_characterizations}

\begin{figure}[t]
\begin{center}
\includegraphics[clip, width=0.49\textwidth]{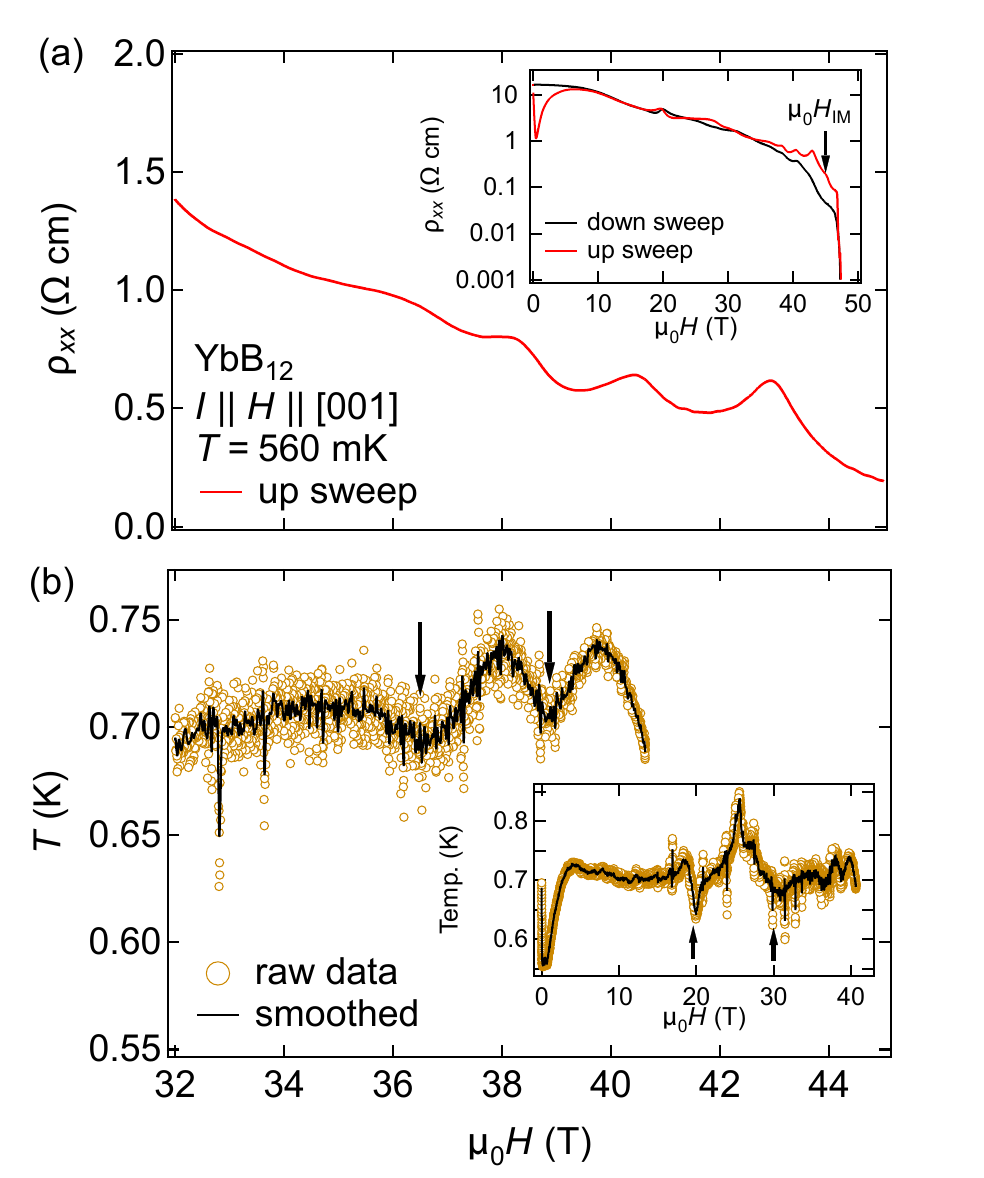}
\end{center}
\caption{
(a) Magnetic-field dependence of the resistivity, $\rho_{xx}$, of YbB$_{12}$ for $\boldsymbol{I} || \boldsymbol{H} || [001]$ at 560 mK during the up sweep of the magnetic field.
The inset shows the data over the full field range with the red (black) line indicating the up (down) sweep.
The arrow indicates the IM transition at about 45 T.
(b) Magnetic-field dependence of the sample temperature during the down sweep of the magnetic field aligned along [001].
The open circles show the raw data, and the solid line indicates 10-point binomial-smoothed data.
The down arrows indicate the fields at 36.5 and 38.9 T, where local minima appear.
The inset shows data down to zero field.
The up arrows indicate local minima at 19.6 and 30 T.
}
\label{Fig_rho_MCE}
\end{figure}

We first performed MR measurements on YbB$_{12}$ to confirm the appearance of previously reported oscillatory features in the insulating phase [Fig.\ \ref{Fig_rho_MCE}(a)].
The resistivity, $\rho_{xx}$, shows a decrease at the IM transition field of $\mu_0 H_\mathrm{IM} \approx 45$ T [inset of Fig.\ \ref{Fig_rho_MCE}(a)].
The hysteresis in $\rho_{xx}$ may be attributed to the MCE
\cite{Terashima_PRL120}.
Between 32 and 44 T, $\rho_{xx}$ shows an oscillating behavior with a MQO frequency of about 700 T, similar to that observed in previous studies in this field range
\cite{Xiang_Science362, Sato_NatPhys15,miz24}, 
indicating MQOs in the insulating phase.

\begin{figure*}[t]
\begin{center}
\includegraphics[clip, width=1.0\textwidth]{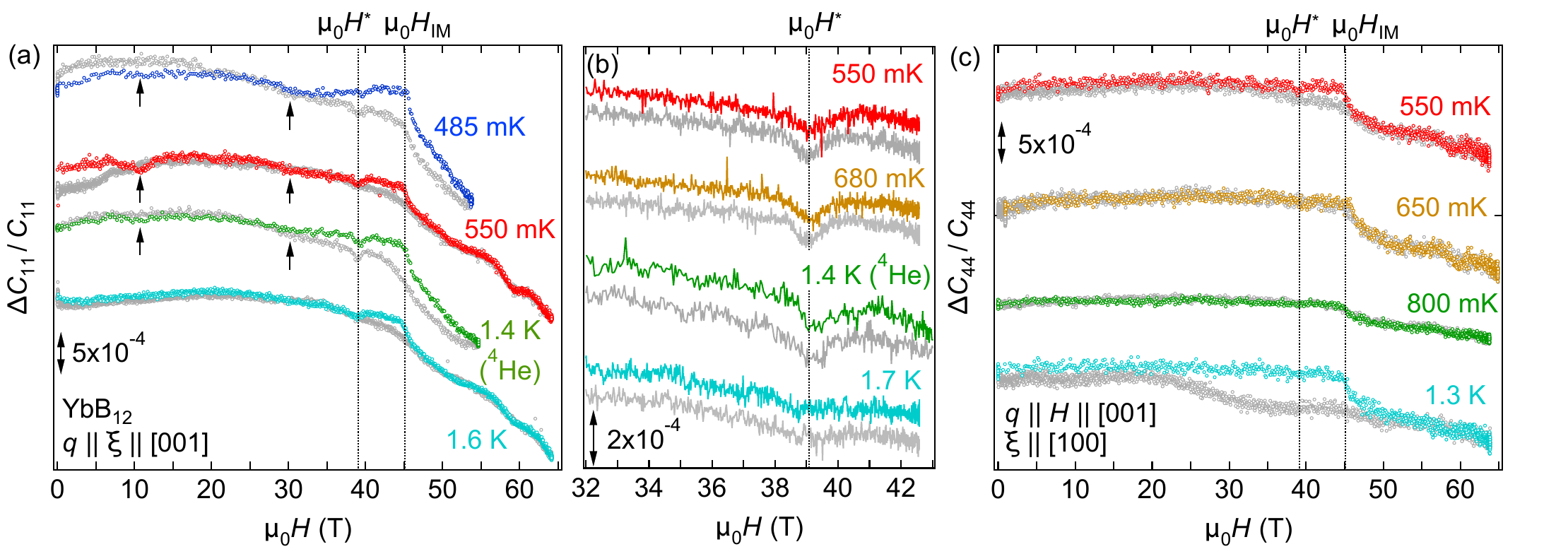}
\end{center}
\caption{
Magnetic-field dependence of the relative change of the longitudinal elastic constant $\Delta C_{11}/C_{11} =  [C_{11}(H) - C_{11}(H = 0)]/C_{11}(H = 0)$ at several temperatures for $\boldsymbol{H} || [001]$ up to about (a) 65 T and (b) 44 T in the range of 32--43 T.
Here, the ultrasound wavevector $\boldsymbol{q}$ and the polarization $\boldsymbol{\xi}$ are parallel to [001].
(c) Relative change of the transverse mode ($q || [001]$, $\xi || [100]$) with the elastic constant $\Delta C_{44}/C_{44} = [C_{44}(H) - C_{44}(H = 0)]/C_{44}(H = 0)$ at several
temperatures for $\boldsymbol{H} || [001]$.
The arrows in panel (a) show weak anomalies around 10 and 30 T.
The data at 485 mK and 1.4 K were measured at the ISSP, and the others at HLD.
The colored (gray) symbols indicate data taken during the up (down) sweep.
The data sets are shifted vertically for clarity.
The vertical dashed lines indicate $\mu_0 H^\star = 39$ T and $\mu_0 H_\mathrm{IM} = 45$ T. 
}
\label{Fig_elastic}
\end{figure*}

We further measured the sample temperature change 
during field sweep in the quasi-adiabatic conditions to study
the MCE [Fig.\ \ref{Fig_rho_MCE}(b)].
We observed some pronounced anomalies with minima at about 20 and 30 T [inset in Fig.\ \ref{Fig_rho_MCE}(b)], which coincide with double-peak structures in the specific heat
\cite{Yang_NatCommun15}.
The anomaly at lower magnetic fields may correspond to the field where the Hall coefficient changes sign or may be related to a Schottky anomaly arising from magnetic defects.
\cite{Xiang_PRX12, Yang_NatCommun15}.
In addition, we observed an oscillatory signal above 36 T with minima at about 36.5 and 38.9 T.
In an earlier study, Chen \textit{et al}.\ reported oscillations in the specific heat with peaks at
about 38.5 and 41.3 T, which correlate with our MCE data \cite{Chen_PRL135}.
The high-field oscillations in our MCE data result in a MQO frequency of about 700 T, which agrees with previously reported values in the insulating state of YbB$_{12}$
\cite{Xiang_Science362, Sato_NatPhys15,miz24,Chen_PRL135}.
Our MR and MCE results, therefore, reflect the high quality of our YbB$_{12}$ sample, which did not deteriorate during its processing for the ultrasonic measurements.

\subsection{High-field ultrasonic measurements}
\label{sub_ultrasound}

Having confirmed the good quality of our YbB$_{12}$ sample, we performed high-magnetic-field ultrasonic measurements to search for MAQOs. 
Figure \ref{Fig_elastic}(a) shows the field dependence of the longitudinal elastic constant $C_{11}$ for $\boldsymbol{H} || [001]$.
We could not resolve MAQOs in the insulating state, but observed several anomalies.
Below $\mu_0 H_\mathrm{IM}$, $\Delta C_{11}/C_{11}$ exhibits weak dip- and kink-like structures around 10 and 30 T, respectively, which are more clearly observed in the up-sweep data.
These anomalies appear at 1.4 K and lower temperatures.
Furthermore, $\Delta C_{11}/C_{11}$ shows another clear dip structure at $\mu_0 H^\star = 39$ T.
Above $\mu_0 H_\mathrm{IM}$, $\Delta C_{11}/C_{11}$ shows a rapid decrease with increasing field, which is consistent with our previous results 
\cite{Kurihara_PRB103}.
The hysteresis found for up and down field sweeps is presumably related to the MCE caused by the IM transition 
\cite{Terashima_PRL120}.

To reduce the influence of irreversible thermal effects in our measurements, we also performed ultrasonic measurements in pulsed fields with a reduced maximum field of less than 44 T below $\mu_0 H_\mathrm{IM}$ [Fig.\ \ref{Fig_elastic}(b)].
In these experiments, we can observe the temperature-independent anomaly at $\mu_0 H^\star$ at all measured temperatures with significantly reduced hysteresis.
However, we could not resolve MAQOs in the insulating state.

Above $\mu_0 H_\mathrm{IM}$, clear MAQOs appear in $\Delta C_{11}/C_{11}$, as visible in the data up to 65 T at 550 mK and 1.6 K [Fig.\ \ref{Fig_elastic}(a)].

To further search for MAQOs in the insulating phase, we also measured the elastic constant $C_{44}$ of a transverse acoustic mode [Fig.\ \ref{Fig_elastic}(c)].
$\Delta C_{44}/C_{44}$ exhibits a clear anomaly at $\mu_0 H_\mathrm{IM}$ and weak MAQOs in the FIM phase at 550 and 650 mK.
However, again, we cannot resolve any MAQOs in the insulating state.
In addition, in this transverse mode, no anomalies appear around 10 and 30 T, and, if at all, we resolve only a very weak feature at $\mu_0 H^\star = 39$ T.

\section{Discussion}
\label{sect_Discussion}

In order to reconcile the anomalies in the elastic constants found in the insulating phase, we compare our results to those of previous studies.
The kink-like feature at 30 T in $C_{11}$ may be related to anomalies found in electrical-transport, torque, and MCE measurements 
\cite{Xiang_Science362, Xiang_PRX12, Yang_NatCommun15}.
Indeed, for the latter, we found a strong minimum in the MCE as well [inset of Fig.\
\ref{Fig_rho_MCE}(b)].
Xiang \textit{et al}.\ proposed a Lifshitz transition of quasiparticle FSs formed by charge-neutral fermions at this and other fields 
\cite{Xiang_Science362, Xiang_PRX12}.
If this Lifshitz transition is indeed responsible for the $C_{11}$ anomaly at 30 T, our results would imply that these FSs couple to the tensile strain $\varepsilon_{xx}$ related to $C_{11}$.  
However, the absence of a feature in $C_{11}$ at the other proposed Lifshitz transition at 19.6 T
\cite{Xiang_PRX12}
remains unclear.
Further, the dip in $C_{11}$ at about 10 T cannot be related to previously reported anomalies in YbB$_{12}$ and needs further studies to be confirmed.

In the FIM phase above $\mu_0 H_\mathrm{IM}$, we resolved MAQOs in $C_{11}$ and $C_{44}$.
From the observed maxima in $C_{11}$ [Fig.\ \ref{Fig_elastic}(a)], we estimate a QO frequency of 620(50) T.
This value is consistent with the low-frequency oscillations reported for the FIM phase with $H || [001]$ in previous studies 
\cite{miz24, Xiang_NatPhys17}, 
in particular in Ref. \cite{Xiang_NatPhys17}, which investigated samples from the same source as the present work.
On the other hand, we could not resolve the higher QO frequencies reported in previous studies
\cite{Liu_npj7}.
Further measurements with different techniques at higher magnetic fields would be needed for a more detailed study of the FS reconstruction in the FIM state.
This, however, is beyond the scope of this work.
Here, we stress that we were able to observe MAQOs in the FIM phase, which are in line with previous reports.

An intriguing result is the observation of the dip structure at $\mu_0 H^\star
= 39$ T in $C_{11}$.
We can exclude that this feature is caused by MAQOs. 
Indeed, for the reported QO frequency of about 700 T
\cite{Xiang_Science362, Sato_NatPhys15,miz24,Chen_PRL135}, 
we would expect two further dips (or oscillations) between $\mu_0 H^\star$ and $\mu_0 H_\mathrm{IM}$.
The origin of the dip structure at 39 T remains elusive.
Experiments using other techniques have not revealed any clear anomaly at this magnetic field.
Further strain-sensitive measurements would be needed to elucidate a possible
field-induced phase transition.

The important question remains why we did not observe MAQOs in the insulating phase of YbB$_{12}$.
Before discussing possible microscopic origins, we first address experimental factors that could obscure MAQOs in the insulating phase. 
In the present study, the MR and ultrasonic measurements were performed using the same pulsed-field setup (see Table \ref{Table_magnet}).
Although the MCE measurements reported previously were performed using a different pulse magnet
\cite{Terashima_PRL120},
the duration of the pulses is comparable to that used in the present study.
These previous studies have shown that the sample temperature decreases near the IM transition, rather than increases. 
Furthermore, clear MAQOs are resolved in the field-induced metallic phase, indicating sufficient sensitivity of the ultrasonic technique. 
Therefore, heating effects inherent to pulsed-field measurements are unlikely to be the primary reason for the absence of detectable MAQOs in the insulating phase.

\begin{figure}[t]
\begin{center}
\includegraphics[clip, width=0.49\textwidth]{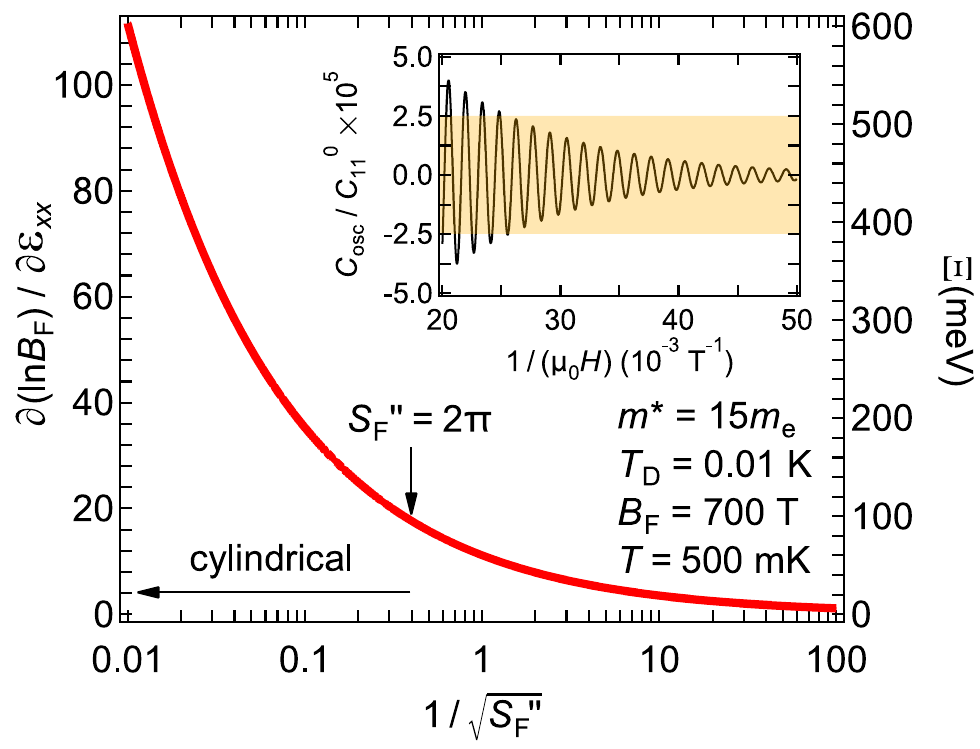}
\end{center}
\caption{
Estimated upper bounds of $\partial \left( \ln B_\mathrm{F} \right) / \partial \varepsilon_{xx}$ and deformation potential $\Xi$ as a function of $1/\sqrt{S_\mathrm{F}^{''}}$.
The horizontal line reflects the degree of two-dimensionality of the Fermi surface, with larger $1/\sqrt{S_\mathrm{F}^{''}}$ corresponding to more cylindrical character. 
The inset shows reproduced MAQOs of YbB$_{12}$ in the range of 20--50 T.
The shaded orange region indicates the experimental resolution $\mathit{\Delta} C/C \sim 5 \times 10^{-5}$.
}
\label{Fig_Upperbounds}
\end{figure}

We now turn to the intrinsic origin of the suppressed MAQOs in the insulating phase by the quasiparticle-phonon coupling.
MAQOs originate from interactions between acoustic phonons with small wavevectors and fermionic quasiparticles on Landau levels at the Fermi energy [Fig.\ \ref{Fig1}(c)].
In the present insulating state, proposed charge-neutral fermions might form Landau levels
\cite{Sato_NatPhys15, Xiang_NatPhys17, Xiang_PRX12,Chen_PRL135}.
If true, contributions of the interaction between these charge-neutral fermionic quasiparticles and phonons to the MAQOs is weak in YbB$_{12}$.
To determine the upper bounds of the interaction energy, we consider the Lifshitz-Kosevich formula for the MAQOs
\cite{Shoenberg_Text, Kataoka_JPSJ62, Luthi_Text, Scindler_PRB102}.
Based on the MQOs observed in the insulating phase, we adopt the cyclotron mass $m^\star \sim 15 m_e$, where $m_e$ is the free electron mass, the Dingle temperature $T_\mathrm{D} = 0.01$ K, and the oscillation frequency $B_\mathrm{F} = 700$ T
\cite{Xiang_Science362}.
Using these parameters, we can reproduce ideal MAQOs in $C_{11}$ at $T = 500$ mK, where the oscillation amplitude at 39 T is set equal to the experimental resolution $\mathit{\Delta} C/C_{11} \sim 5 \times 10^{-5}$
(see the inset in Fig. \ref{Fig_Upperbounds}).
From this analysis, we obtain the relationship between $\partial B_\mathrm{F} / \partial \varepsilon_{xx} \propto \partial S_\mathrm{F} / \partial \varepsilon_{xx}$ and $S_\mathrm{F}^{''}$ in Eq. (\ref{dB_Fd_eps}).
Here $\varepsilon_{xx}$ is the tensile strain associated with the elastic constant $C_{11}$, $\partial S_\mathrm{F} / \partial \varepsilon_{xx}$ represents the strain-induced change in the cyclotron orbit, and $S_\mathrm{F}^{''}$ represents the curvature of the extremal Fermi surface cross-section $S_\mathrm{F}$ perpendicular to the magnetic field.
Using this relationship, we determine the upper bounds of $\partial \left( \ln B_\mathrm{F} \right) / \partial \varepsilon_{xx} = \partial \left( \ln S_\mathrm{F} \right) / \partial \varepsilon_{xx}$ and the deformation potential $\Xi = \hbar \partial \left( e B_\mathrm{F} / m^\star \right) /  \partial \varepsilon_{xx} $, which characterize the coupling between the fermionic quasiparticles and acoustic phonons, as shown in Fig. \ref{Fig_Upperbounds}.

To obtain quantitative estimates, we consider realistic parameters for YbB$_{12}$.
Based on the reported angle dependence of the MQOs
\cite{Xiang_Science362},
we consider that a nearly isotropic case with $S_\mathrm{F}^{''} \sim 2 \pi$, corresponding to a spherical Fermi surface, provides a reasonable reference for YbB$_{12}$.
In this case, we obtain $\partial \left( \ln B_\mathrm{F} \right) / \partial \varepsilon_{xx} = \partial \left( \ln S_\mathrm{F} \right) / \partial \varepsilon_{xx} \approx 18$, which is comparable to the values reported in other materials, whereas $\Xi \approx 95$  meV is smaller than values reported previously
\cite{Walther_PR174, Tekippe_PRB6, Vurgaftman_JAP89, Settai_JPSJ61, Settai_JPSJ63, Matsui_JPSJ64}.
This reduction mainly originates from the large cyclotron mass, which lowers the energy scale associated with the cyclotron motion.
As a result, the amplitude of MAQOs is strongly suppressed and can fall below the experimental resolution.
Since these values are derived as upper bounds constrained by the absence of detectable MAQOs, the actual interaction strength in the insulating phase is likely smaller. 
To directly determine the interaction strength, QO measurements under uniaxial strain
\cite{Schindler_PRB102}
would be necessary.
We note that, considering the Fermi wavenumber, relaxation time, Fermi velocity, and mean free path, the conditions for ultrasonic waves to probe the quasiparticles are well fulfilled, indicating that the absence of MAQOs is not due to a kinematic limitation (see Appendix \ref{Appendix A}).

Although a weak contribution of the quasiparticle-phonon interaction to the cyclotron motion is indicated, we observed a kink-like feature in $C_{11}$ at 30 T, which is attributed to the field-induced Lifshitz transition, as discussed above.
This suggests that the transition can still occur and manifest itself through a non-oscillatory change in the electronic structure, even when quantum oscillations are not resolved.
Such behavior can be explained by the large cyclotron mass, which reduces the relevant energy scale of the cyclotron motion and, thereby, diminishes the oscillatory response, while leaving the underlying change in the electronic structure relatively unaffected.
To deepen the understanding of the unconventional features in the insulating phase of YbB$_{12}$, measurements with considerably higher resolution or other strain-related techniques, such as magnetostriction measurements 
\cite{Ikeda_RSI88, Miyake_RSI91},
might be needed.

\section{Conclusions}
\label{sect_Conclusion}
In this work, we searched for magnetoacoustic quantum oscillations in the insulating phase of the Kondo insulator YbB$_{12}$. 
We investigated a single crystal that, in previous investigations, showed quantum oscillations in the magnetoresistance and magnetic torque in the insulating phase \cite{Xiang_Science362, Sato_NatPhys15}.
We observed such oscillations in magnetoresistance and magnetocaloric effect measurements as well.
However, we could not resolve any magnetoacoustic quantum oscillations in the insulating state. 
One reason for that might be a weak effective coupling between the proposed fermionic quasiparticles and acoustic phonons in YbB$_{12}$, which is suppressed in the oscillatory response due to the large cyclotron mass.
From the absence of detectable oscillations, we determined upper bounds on $\partial\!\left( \ln B_\mathrm{F} \right)\!/\partial\varepsilon$ and the corresponding deformation potential, providing quantitative constraints on the quasiparticle–phonon coupling in the insulating phase.
On the other hand, in the metallic state, above the field-induced insulator-metal transition at about 45 T, the coupling between charged quasiparticles and phonons was sufficiently strong to observe clear quantum oscillations in two different acoustic modes.
The origin of several weak anomalies in the longitudinal elastic constant $C_{11}$ in the insulating state remains unclear. 
Further work is needed to understand the nature of the magnetic quantum oscillations observed in Kondo insulators.

\section*{Acknowledgment}

The authors thank Yuki Sato, Zhuo Yang, and Tatsuya Yanagisawa for helpful discussions.
This work was supported in part by JSPS Bilateral Joint Research Projects (JP JSBP120193507), Grants-in-Aid for Early-Carrier Scientists (KAKENHI JP 20K14404), Transformative Research Areas (A) (JP 23H04862, 24H01629), Fund for the Promotion of Joint International
Research [Fostering Joint International Research (B)] (JP 21KK0046), and JSPS Overseas Challenge Program for Young Researchers.
The authors also acknowledge the support of the HLD at HZDR, member of the European Magnetic Field Laboratory (EMFL), the Deutsche Forschungsgemeinschaft (DFG) through the W\"urzburg-Dresden Cluster of Excellence on Complexity, Topology and Dynamics in Quantum Matter---$ctd.qmat$ (EXC 2147, project No.\ 390858490), and the BMBF via DAAD (project No.\ 57457940).

\appendix
\setcounter{figure}{0}
\renewcommand{\thefigure}{A\arabic{figure}}

\section{
Lifshitz-Kosevich formula for magnetoacoustic quantum oscillations
}
\label{Appendix A}

\begin{widetext}
Here we present the MAQOs based on the Lifshitz-Kosevich formula.
Starting from the free energy given by
\cite{Shoenberg_Text, Kataoka_JPSJ62, Luthi_Text, Scindler_PRB102}
\begin{align}
\label{LK free energy}
\mathit{\Omega}
=\left( \frac{e}{2 \pi \hbar} \right)^{3/2} \left( \frac{e \hbar}{m^\star}\right) \frac{1}{\pi^2 \sqrt{S_\mathrm{F}^{''} } } 
    \sum_p \left( \frac{B}{p} \right)^{5/2} R_T^p R_\mathrm{D}^p R_\mathrm{S}^p 
        \cos\left[ 2 \pi p \left( \frac{B_\mathrm{F}}{B} - \frac{1}{2} \right) \pm \frac{\pi}{4} \right]
,
\end{align}
we can describe the oscillatory contribution to the elastic constant, $C_\mathrm{osc} = \partial^2 \mathit{\Omega} / \partial^2 \varepsilon$, as 
\begin{align}
C_\mathrm{osc}
= - 4\left( \frac{e}{2 \pi \hbar} \right)^{3/2} \left( \frac{e \hbar}{m^\star}\right) \frac{1}{ \sqrt{S_\mathrm{F}^{''} } } 
        \left( \frac{\partial B_\mathrm{F}}{\partial \varepsilon} \right)^2 
            \sum_p \left( \frac{B}{p} \right)^{1/2} R_T^p R_\mathrm{D}^p R_\mathrm{S}^p 
                \cos\left[ 2 \pi p \left( \frac{B_\mathrm{F}}{B} - \frac{1}{2} \right) \pm \frac{\pi}{4} \right]
.
\end{align}
\end{widetext}
Here, $e$, $\hbar$, $m^\star$, $S_\mathrm{F}^{''}$, and $B_\mathrm{F} = \hbar S_\mathrm{F} / \left( 2\pi e \right)$ denote the elementary charge, the Dirac constant, the cyclotron mass, the curvature of the extremal Fermi surface cross-section $S_\mathrm{F}$ perpendicular to the magnetic field, and the oscillation frequency, respectively.
The damping factors $R_T^p$, $R_\mathrm{D}^p$, and $R_\mathrm{S}^p$ are given by 
\begin{align}
\label{R_T}
R_T^p
&= \left( \lambda \frac{m^\star}{m_e} \frac{T}{B} p \right) / \sinh\left( \lambda \frac{m^\star}{m_e} \frac{T}{B} p  \right)
,
\end{align}
\begin{align}
\label{R_D}
R_\mathrm{D}^p
&= \exp\left[- \lambda \frac{m^\star}{m_e} \frac{T_\mathrm{D}}{B} p \right] 
, 
\end{align}
\begin{align}
\label{R_S}
R_\mathrm{S}^p 
&= \cos\left( p \pi \frac{g m^\star}{2m_e} \right)
,
\end{align}
\begin{align}
\lambda 
&= \frac{2 \pi^2 k_\mathrm{B} m_e} {e \hbar}
,
\end{align}
where $m_e$, $T_\mathrm{D}$, $g$, and $k_\mathrm{B}$ are the free electron mass, the Dingle temperature, the effective $g$-factor, and the Boltzmann constant, respectively.
Here, we assumed $g = 2$.
As shown in the inset in Fig. \ref{Fig_Upperbounds}, using $m^\star = 15 m_e$, $T_\mathrm{D} = 0.01$ K, $B_\mathrm{F} = 700$ T reported in Ref. \cite{Xiang_Science362}, and $C_{11} = 36.0 \times 10^{10}$ J/m$^3$ at 500 mK in Ref. \cite{Kurihara_PRB103},
we can reproduce ideal MAQOs in the elastic constant $C_{11}$ at $T = 500$ mK.
The oscillation amplitude around 39 T is set to match the experimental resolution, $\mathit{\Delta}C / C = C_\mathrm{osc} / C \approx 5 \times 10^{-5} $, $C_{11}$.
From this condition, we obtain 
\begin{align}
\label{dB_Fd_eps}
\frac{1}{ \sqrt{S_\mathrm{F}^{''} } } 
        \left( \frac{\partial B_\mathrm{F}}{\partial \varepsilon_{xx} } \right)^2 
\approx 6.1 \times 10^7 \ \mathrm{T^2}
.
\end{align}
Using this relationship, we determine the upper bounds of $\partial\!\left( \ln B_\mathrm{F} \right)\!/ \partial \varepsilon_{xx} = \partial\! \left( \ln S_\mathrm{F} \right) \!/ \partial \varepsilon_{xx}$ and the corresponding deformation potential,
\begin{align}
\label{deformation potential}
\Xi =  \hbar \frac{\partial  }{\partial \varepsilon_{xx}} 
    \left( \frac{ e B_\mathrm{F}}{m^\star} \right)
,
\end{align}
which characterizes the strain-induced change in the cyclotron energy,
as shown in Fig. \ref{Fig_Upperbounds}.

Based on the Lifshitz-Kosevich formula, we can estimate the relaxation time derived from the Dingle temperature $\tau_\mathrm{D} = 1.2 \times 10^{-10}$ s using $\tau_\mathrm{D} = \hbar / \left( 2 \pi k_\mathrm{B} T_\mathrm{D} \right)$, the Fermi wavenumber $k_\mathrm{F} = 0.15$ \AA$^{-1}$\ from $S_\mathrm{F} = \pi k_\mathrm{F}^2$, the Fermi velocity $v_\mathrm{F} = 1.1 \times 10^4$ m/s from $v_\mathrm{F} = \hbar k_\mathrm{F} / m^\star$, and the mean free path $l_\mathrm{mean} = 1.4 \times 10^{-6}$ m using $l_\mathrm{mean} = v_\mathrm{F} \tau_\mathrm{D}$.
Based on the ultrasonic frequency $f \sim 1.0 \times 10^8$ Hz and the sound velocity $v_{11} \sim 8.6 \times 10^3$ m/s, we can estimate an acoustic wavelength $\lambda \sim 86$ $\mathrm{\mu m}$ and corresponding wavevector $q \sim 7.3 \times 10^4$ m$^{-1}$.
Comparing this with $k_\mathrm{F}$, we find $q_{11} \ll k_\mathrm{F}$, indicating that the ultrasonic waves probe the long-wavelength limit relevant for quasiparticles.
In addition, using the estimated $\tau_\mathrm{D}$, we obtain $ 2 \pi f \tau_\mathrm{D} \ll 1$, confirming that the quasiparticles can adiabatically follow the ultrasonic excitation.
Furthermore, $v_\mathrm{F}$ is comparable to $v_{11}$, indicating no significant kinematic mismatch between quasiparticles and phonons.
These considerations demonstrate that the conditions for ultrasonic waves to probe fermionic quasiparticles are well satisfied.


\bibliography{Reference}

\end{document}